\documentstyle[sprocl]{article}

\def\PLB{{\em Phys. Lett.}  B}

\def\PRD{{\em Phys. Rev.} D}

\def\bea{\begin{eqnarray}}
\def\eea{\end{eqnarray}}
\def\be{\begin{eqnarray}}
\def\ee{\end{eqnarray}}

\newcommand{\nc}{\newcommand}
\newcommand{\bi}{\bibitem}
\newcommand{\om}{\omega}
\newcommand{\ds}{\partial \!  \! \! /}
\newcommand{\Zs}{Z \! \! \! \! /}
\newcommand{\Ws}{W \! \! \! \! \! /}
\newcommand{\rar}{\rightarrow}

\newcommand{\lrar}{\leftrightarrow}
\newcommand{\dm}{\delta m^2}

\newcommand{\num}{\nu_\mu}
\newcommand{\nue}{\nu_e}
\newcommand{\nut}{\nu_\tau}

\newcommand{\cost}{\cos \theta}
\newcommand{\sint}{\sin \theta}
\newcommand{\stw}{\sin 2 \theta}
\newcommand{\ctw}{\cos 2 \theta}
\begin{document}
\title{NEUTRINO OSCILLATIONS AND COSMOLOGY } 
\author{ A. D. DOLGOV  }
\address{Teoretisk Astrofysik Center, Juliane Maries Vej 30,\\
DK-2100, Copenhagen, Denmark; \\
ITEP, 117259, Moscow, Russia\\
and \\
INFN, sezione di Ferrara Via Paradiso, 12 - 44100 \\
Ferrara, Italy}
 
\maketitle\abstracts{
Phenomenology of neutrino oscillations in vacuum and in cosmological 
plasma is considered. Neutrino oscillations in vacuum are usually 
described in plane wave approximation. In this formalism there is an 
ambiguity if one should assume $\delta p =0$ and correspondingly 
$\delta E\neq 0$ or vice versa $\delta E =0$ and  $\delta p \neq 0$, 
or some other condition. We will use the standard set of quantum field 
theory rules and show that the theoretical description is unambiguous 
and that the state of oscillating neutrinos is determined by the 
conditions of their registration. The wave packet formalism is a natural 
result of such an approach.\\
Oscillating neutrinos in cosmological plasma cannot be described in 
terms of wave function because of a fast loss of coherence due to elastic 
or inelastic scattering, so that one should use the density matrix  
formalism. Kinetic equations for the density matrix of oscillating 
neutrinos are derived. Numerical and semi-analytical solutions of the 
equations are examined.  In particular, a possibility of amplification 
of cosmological lepton asymmetry in the sector of active neutrinos mixed 
to sterile ones is critically discussed. 
}

\section{Neutrino oscillations in vacuum. Basic concepts.}

As is very well known, neutrinos possibly oscillate if (and
because) their mass eigenstates are different from their interaction
eigenstates~\cite{pontecorvo57}$^-$\cite{bahcall69}
(for a review and more references see~\cite{bilenky78}).
In other words, the mass matrix of different neutrino species
is not diagonal in the basis of neutrino flavors: $ [\nue,\,\num,\,\nut]$.
The latter is determined by the interaction with charged leptons, so that
a beam of e.g. electrons would produce $\nue$ which is a mixture of
several different mass states. An important condition is that the masses 
are different, otherwise oscillations would be unobservable. Indeed,
if all the masses are equal, the mass matrix would be proportional to the unit 
matrix which is diagonal in any basis. 

Of course not only neutrinos are capable to oscillate. All particles
that are produced in the same reactions will do that, but usually the
oscillation frequency, $\omega_{osc} \sim \delta m^2 /2E$ is so huge
and correspondingly the oscillation length
\be
l_{osc}= 2 p /\delta m^2\,
\label{losc}
\ee
is so small, that
the effect is very difficult to observe. Here $E$ and $p$ are respectively
the energy and momentum of the particles in consideration 
and $\delta m^2 = m_1^2 - m_2^2$. The expression
is written in relativistic limit. Only for $K$-mesons and hopefully for
neutrinos the mass difference is so small that $l_{osc}$ is, or may be,
macroscopically large.

The neutrino Lagrangian can be written as follows:
\be
{\cal L_\nu} = i \bar \nu \ds \nu + \bar \nu {\cal M} \nu
+ \bar \nu \Zs \nu + \bar \nu \Ws \,\,l
\label{lnu}
\ee
where the vector-column $\nu = [\nue,\,\num,\,\nut]^T$ is the operator of
neutrino field in interaction basis, $l = [ e,\, \mu,\,\tau]^T$ is the
vector of charged lepton operators; the last two terms describe 
respectively  neutral and charge current interactions (with $Z$ and $W$
bosons). The upper index ``T'' means transposition.
The matrix ${\cal M}$ is the mass matrix and by assumption it is 
non-diagonal in the interaction basis. This is not necessary but quite
natural because normally masses know nothing about interactions.

Transformation between mass and interaction eigenstates is realized by an
orthogonal matrix $U$ with the entries that are parameters of the theory
to be determined by experiment. In the simplest case of only
two mixed particles the matrix $U$ has the form
\bea
U= \left( \begin{array}{cc} \cost & \sint \\
-\sint & \cost \end{array} \right)
\label{u2}
\eea
If for example the only noticeable mixing is between electronic and muonic
neutrinos, then the flavor eigenstates are related to the mass eigenstates
$\nu_{1,2}$ as:
\bea
\nue & =& \nu_1\, \cost +  \nu_2\, \sint\,, \nonumber \\
\num &=& - \nu_1\, \sint + \nu_2\, \cost 
\label{nuemu}
\eea
Thus if electronic neutrinos are produced on a target by a beam of
electrons, their propagating wave function would have the form
\be
\psi_{\nue} (\vec r, t) = \cost \,|\nu_1\rangle\, e^{ik_1 x} +
\sint\, |\nu_2\rangle\, e^{ik_2 x}
\label{psinue}
\ee
where $kx = \om t - \vec k \vec r$ and sub-$\nue$ means that the initial
state was pure electronic neutrino. Below (in this section only)
we denote neutrino energy by $\om$
to distinguish it from the energies of heavy particles that are denoted as
$E$. We assume, as is normally done, plane wave representation of the wave 
function.

If such a state hits a target, what is the probability of producing an
electron or a muon? This probability is determined by the fraction of
$\nue$ and $\num$ components in the wave function $\psi_{\nue}$ at 
space-time point $x$. The latter can be found by re-decomposition of
$\nu_{1,2}$ in terms of $\nu_{e,\mu}$:
\be
\psi_{\nue} (\vec r, t) = \cost\,e^{ip_1x}\, \left( \cost\, |\nue \rangle -
\sint\, |\num \rangle \right)+
\sint\, e^{ip_2 x}\, \left( \sint\, |\nue \rangle +
\cost\, |\num \rangle \right)
\label{psinuemu}
\ee
One can easily find from that expression that the probability to register
$\nue$ or $\num$ are respectively:
\bea
P_{\nue}(\vec r, t) \sim 
\cos^4 \theta +\sin^4\theta + 2\sin^2 \theta\, \cos^2\theta \cos\, 
\delta \Phi\,,
\label{pnue}
\eea
\bea
P_{\num}(\vec r, t) \sim 2 \sin^2\theta\, \cos^2\theta \left( 1 -
\cos\, \delta \Phi \right)
\label{pnum}
\eea
Here $\delta \om = \om_1 -\om_2$, $\delta \vec k = \vec k_1 - \vec k_2$,
and 
\be
\delta \Phi\, (\vec r, t) =  \delta \om\, t - \vec \delta k\, \vec r 
\label{deltaphi}
\ee
The energy difference between the mass eigenstates is
\be
\delta \om= {\partial\om \over \partial m^2}\, \delta m^2 +
{\partial \om \over \partial \vec k}\, \delta \vec k
\label{deltae}
\ee
Using this expression we find for the phase difference
\be
\delta \Phi\, (\vec r, t) = {\delta m^2 \over 2\om }\, t + \delta \vec k 
\left( {\vec k \over \om}\, t - \vec r \right)
\label{deltaphi2}
\ee

The standard result of the neutrino oscillation theory are obtained if
one assumes that: 1) $\delta \vec k =0$, 2) $\vec k = \om \vec v$, and
3) $t=r/v$:
\be
\delta \Phi = {\delta m^2 r \over 2 k}
\label{deltaphist}
\ee
Each of these assumptions is difficult to understand and, even more, some 
of them, in particular, $\delta k=0$ may be explicitly incorrect
(see below). Both the second and the third conditions are fulfilled for a 
classical motion of a point-like body, however their validity should be
questioned for a quantum mechanical particle (for a wave). 
Despite all that the final result (\ref{deltaphist})
is true and if there are some corrections, they can be trivially understood.

Basic features of neutrino oscillations were discussed in many papers. An
incomplete list of references includes~\cite{nussinov76}$^-$\cite{nauenberg99}.
One can find more citations and discussion in the above quoted papers and 
in the books~\cite{bahcall89}$^-$\cite{fukugita94}.
Still some confusion and indications on possible controversies reappear 
from time to time in the literature, 
so it seems worthwhile to present a consistent derivation
of eq.~(\ref{deltaphist}) from the first principles.
A large part of this section is based on the discussions (and unpublished 
work) with A.Yu. Morozov, L.B. Okun, and M.G. Schepkin. 

Let us consider a localized source that produces oscillating neutrinos;
we keep in mind, for example, a pion decaying through the channel
$\pi \rar \mu + \nu_\mu$. The wave function of the source $\psi_s (\vec r,t)$
can be Fourier decomposed in terms of plane waves:
\bea
\psi_s (\vec r,t) = \int d^3 p\,\, C(\vec p -\vec p_0) 
e^{iEt -i \vec p \vec r}&\approx& 
\nonumber \\
e^{iE_0 t - i\vec p_0 \vec r} 
\int d^3 q\, C(\vec q) \exp \left[ -i\vec q \left(\vec r -
 \vec V_0 t \right)\right]
&=& e^{iE_0 t -i\vec p_0 \vec r}\, \tilde C \left( \vec r - \vec V_0 t \right)
\label{psis}
\eea
where $\vec V_0 = \vec p_0 /E_0$.
It is the standard wave packet representation. The function 
$C(\vec p -\vec p_0)$ is assumed to be sharply peaked around the central 
momentum $\vec p_0$ with dispersion $\Delta \vec p$. The particle is, by 
construction, on-mass-shell, i.e. $E^2 = p^2 + m^2$. This is of course also 
true for the central values $E_0$ and $p_0$.
As the last expression shows, the particle
behaves as a plane wave with frequency and wave vector given respectively
by $E_0$ and $\vec p_0$ and with the shape function (envelope) given
by $\tilde C (\vec r- \vec V_0 t)$ which is 
the Fourier transform of $C(\vec q)$. Evidently the envelope moves 
with the classical velocity  $\vec V_0 $. Characteristic
size of the wave packet is $l_{pack} \sim 1/\Delta p$.

Let us consider the pion decay, $\pi \rar \mu + \nu$. One would naturally
expect $\delta \om \sim \delta k \sim \delta m^2 /E$. If this is true
the probability of oscillation would be
\be
P_{osc} \sim \cos \left[ { x + b\, \left(x -Vt\right) \over l_{osc}}\right],
\label{posc}
\ee
where $l_{osc}$ is given by expression (\ref{losc}) and
$b$ is a numerical coefficient relating $\delta p $ with $l_{osc}$.
For simplicity the one-dimensional expression is presented.

Thus to obtain the probability of neutrino registration one should average
the factor $(x-Vt)$ over the size of the wave packet and for large 
packets, if $b$ is non-negligible,
a considerable suppression of oscillations should be expected.  
The size of the neutrino wave packet from the pion
decay at rest is macroscopically large, 
$l_{pack} \approx c\, \tau_\pi \approx 7.8$ m, 
where $c$ is the speed of light and $\tau_\pi = 2.6 \cdot 10^{-8}$ sec
is the pion life-time. The oscillation length is 
$l_{osc} = 0.4\, {\rm m}\, ( E/ {\rm MeV} / (\delta m^2 /{\rm eV}^2) $,
so $l_{osc}$ could be smaller or comparable to $l_{pack}$ and the effect
of suppression of oscillations due to a finite size of the wave packet 
might be significant. It is indeed true but only for the decay of a moving 
pion, and this suppression is related to 
an uncertainty in the position of decay. To see that we have to abandon
the naive approach described above 
and to work formally using the standard set of quantum
mechanical rules.

Let us assume that neutrinos are produced by a source with the
wave function $\psi_s(\vec x, t)$. This source produces neutrinos together
with some other particles. 
We assume first the following experimental conditions:
neutrinos are detected at space-time point $\vec x_\nu, t_\nu$, while
the accompanying particles are not registered. The complete set of 
stationary states of these particles is given by the wave functions 
$\psi_n \sim \exp (iE_n t)$. The amplitude of registration of propagating
state of neutrino of type $j$ (mass eigenstate) accompanying by other
particles in the state $\psi_n$ is given by
\be
A_j^{(n)} = \int d\vec r_s\, dt_s \psi_s (\vec r_s, t_s)
\psi_n (\vec r_s, t_s) G_{\nu_j} \left(\vec r_\nu - \vec r_s, t_\nu - t_s
\right)
\label{ajn}
\ee
In principle one even does 
not need to know the concrete form of $\psi_n$, the only necessary
property of these functions is the condition that they form a complete set:
\be
\sum_n \psi_n \left(\vec r, t\right) \psi^*_n \left(\vec r\,',t \right)
= \delta \left(\vec r - \vec r\,' \right)
\label{sumn}
\ee
However in what follows we will use for simplicity the eigenfunctions of
momentum, $\psi_n \sim \exp (i \vec p\,\vec r -iEt)$.
 
For the subsequent calculations we need the following representation of the
Green's function which is obtained by the sequence of integration:
\bea
G(\vec r, t) &=& \int {d^4 p_4 \over p^2_4 - m^2} e^{ip_4 x} =
\nonumber \\
&&2\pi \int_{-\infty}^{+\infty} d\omega e^{i\omega t} \int_0^{+\infty} 
{dp p^2 \over \omega^2 - p^2 - m^2} \int_{-1}^1 d\zeta
e^{-i p r \zeta} = 
\nonumber \\
&&{i\pi \over r}  \int_{-\infty}^{+\infty} d\omega e^{-i\omega t} 
\int_{-\infty}^{+\infty} 
{dp p^2 \over \omega^2 - p^2 -m^2 } \left( e^{ipr} - e^{-ipr} \right)
\label{green1}
\eea
Here $\omega$ and $p$ are respectively the fourth and space components of 
the 4-vector $p_4$. We have omitted spin matrices because the final
result is essentially independent of that. The integration over $dp$ was
extended over the whole axis (from $-\infty$ to $+\infty$) because the 
integrand is an even function of $p$. This permit to calculate this integral 
by taking residues in the poles on mass shell: 
$p = \pm \sqrt { \omega ^2 - m^2 +i\epsilon }$. Both poles give the same 
contribution, so
skipping unnecessary numerical coefficients, we finally obtain:
\be 
G( \vec r , t) = {1\over r} \int_{-\infty}^{+\infty} 
d\om e^{-i\om t +i \sqrt{ \om^2 - m^2}\,r }
\label{green2}
\ee
As a source function $\psi_s$ we will take essentially 
expression~(\ref{psis}) but assume that the source is a decaying particle
with the decay width $\gamma$, 
that was born at the moment $t=0$. It corresponds to multiplication of
$\psi_s$ by $\theta (t) \exp (-\gamma t)$. Taking all together we obtain
the following expression for the amplitude:
\bea 
A_j^{(n)}(\vec r, t ) &=& \int_0^{\infty} dt_s 
\int {d\vec r_s \over |\vec r - \vec r_s|} 
\int d\vec p\, C\left( \vec p -
\vec p_0 \right) e^{i E t_s - i \vec p \vec r_s -\gamma t/2}\,
e^{i E_n t_s +i \vec p_n \vec r_s } \nonumber \\
&&\int_{-\infty}^{+\infty} d\om_j e^{i\om_j (t-t_s) -ik_j |\vec r -
\vec r_s | }
\label{ajn2}
\eea
Integrals over $d\vec r_s$ and $d\vec p$ are taken over all infinite space.  
It is worthwhile to remind here that all the momenta are on mass shell,
$E^2 = p^2 +m^2_\pi$ (we assumed that the source is a decaying pion) and
$\omega_j^2 = k_j^2 + m_j^2$, where $m_j$ is the mass of $j$-th neutrino
mass eigenstate.  

The integration over $dt_s$ is trivial and gives the factor 
$\left( E-E_n -\om_j +i\gamma /2 \right)^{-1}$.
The integration over $d\vec r_s$ can be easily done if the registration 
point is far from the source. In this case it is accurate enough to 
take $1/|\vec r - \vec r_s| \approx 1/r$,
while the same quantity in the exponent should be expanded up to the first 
order:
\be
|\vec r -\vec r_s| \approx r - \vec \xi\, \vec r_s
\label{r-rs}
\ee
where $\vec \xi = \vec r /r$ is a unit vector directed from the center of
the source taken at the initial moment $t=0$ to the 
detector at the point $\vec r$.
In this limit the integral over $d\vec r_s$ gives $\delta \left(\vec p -
\vec p_n - \vec k_j \right)$, ensuring momentum conservation:
\be
\vec p = \vec p_n + \vec k_j \equiv \vec p_{\pi,j}
\label{pcons}
\ee
The vector of neutrino momentum is formally defined as
\be
\vec k_j = \vec \xi \, k_j = \vec \xi \sqrt{\om_j^2 - m_j^2}
\label{kj}
\ee 
Ultimately we are left with the integral:
\be
A_j^{(n)} = {1\over r} \int_{-\infty}^{+\infty} d\om_j C\left( \vec p_n
+ \vec k_j - \vec p_0 \right) {e^{i\om_j t -ik_j r} 
\over E_{\pi,j}-E_n -\om_j + i\gamma/2}
\label{ajn3}
\ee
where $E_{\pi,j} = \sqrt{(\vec p_n + \vec k_j )^2 + m_\pi^2 }$.
This integral can be taken in the 'pole approximation' and to do that
we need to expand the integrand near the energy conservation law,
see below eq.~(\ref{e0}),  as follows. 
The neutrino energy is presented as $\om_j = \om_j^{(0)} + \Delta \om_j$.
To avoid confusion
one should distinguish between the deviation of neutrino energy from the
central value given by the conservation law, $\Delta \om_j$, from the 
difference of energies of different neutrino mass eigenstates,
$\delta \om = \om_1 - \om_2$. 
The neutrino momentum is expanded up to the first order in $\Delta \om_j$:
\be
 k_j = \sqrt {\om_j^2 -m^2_j} \approx k_j^{(0)} + 
\Delta \om_j /V_j^{(\nu)}
\label{kjexp}
\ee
where $V_j^{(\nu)} = k_j^{(0)} /\om_j^{(0)}$ is the velocity 
of $j$-th neutrino.
The pion energy is determined by the momentum conservation~(\ref{pcons})
and is given by
\be
E_{\pi,j} = 
\sqrt{ \left(\vec p_n + \vec k_j^{(0)} + \vec\Delta k_j \right)^2 +
m^2_\pi } \approx E^{(0)}_{\pi,j} + \vec V_{\pi,j} \vec \xi\, \Delta \om_j
\label{epi}
\ee  
where the pion velocity is 
$\vec V^{(\pi)}_j = (\vec p_n + \vec k_j^{(0)})/E^{(0)}_{\pi,j}$. 
The neutrino energy, $\om_j^{(0)}$, satisfying the conservation law
is defined from the equation:
\be
E^{(0)}_{\pi,j} - E_n - \om_j^{(0)} =0
\label{e0}
\ee
Now the integral over $\om_j$ is reduced to
\bea
A_j^{(n)} = {e^{i\om_j^{(0)} t - i k_j^{(0)} r }  \over r}
C\left( \vec p_n  + \vec k_j^{(0)} - \vec p_0 \right) 
 \nonumber \\
\int_{-\infty}^{+\infty} d\Delta \om_j\,\,
{e^{i\Delta \om_j \left( t - r/V^\nu_j \right)} \over
\left( \vec V_j^{(\pi)} \vec \xi / V_j^{(\nu)}\right) \Delta \om_j -
\Delta \om_j +i\gamma/2 }
\label{ajn4} 
\eea
The last integral vanishes if $t< V_j^{(\nu)} r$, while in the opposite
case it can be taken as the residue in the pole and we finally
obtain:
\be
A_j^{(n)} = {C\left( \vec p_{\pi,j} - \vec p_0 \right) \over r}\,
\theta \left (r - V_j^{(\nu)} t\right)
\exp  \left(i\om^{(0)} t -i k_j^{(0)}r 
-{\gamma\over 2}{ V_j^{(\nu)} t - r \over V_j^{(\nu)} - 
\vec V^{(\pi)}_j \vec \xi} \right)
\label{ajnfin}
\ee
We obtained a neutrino wave packet moving with the velocity $V^{(\nu)}_j$
with a well defined front (given by the
theta-function) and decaying with time in accordance with the decay law of
the source. The similar wave packet, but moving with a slightly different
velocity describes another oscillating state $\nu_i$. It is evident from
this expressions that phenomenon of coherent oscillations takes place only
if the packets overlap, as was noticed long 
ago~\cite{nussinov76,dolgov81,kayser81}.

The probability of registration of oscillating neutrinos at the space-time
point $(\vec r, t)$ is given by the density matrix
\be
\rho_{ij} = \int d\vec p_n\, A^{(n)}_i \left(\vec r_\nu, t_\nu \right) 
A^{*(n)}_j \left(\vec r_\nu, t_\nu \right)
\label{rhoij}
\ee
The oscillating part of the probability is determined by the phase 
difference~(\ref{deltaphi}) but now the quantities $\delta \om$ and
$\delta k$ are unambiguously defined. To this end we will use the conservation
laws~(\ref{pcons},\ref{e0}). They give:
\be 
\delta \om^{(0)} = \delta E_{\pi}\,\,\, {\rm and}\,\,\, 
\delta \vec k^{(0)} = \delta \vec p_{\pi}
\label{deltaomk}
\ee
The variation of neutrino energy is given by
\be 
\delta \om = V^{(\nu)} \delta k + \delta m^2 / 2 \om
\label{deltaom} 
\ee
while the variation of the pion energy can be found from expression 
(\ref{epi}):
\be
\delta E^{(\pi)} = \vec V^{(\pi)} \delta \vec k 
\label{deltae1}
\ee
From these equations follows
\be
\delta \om = - {\delta m^2 \over 2\om} {\vec V^{(\pi)} \vec \xi
\over V^{(\nu)} - \vec V^{(\pi)} \vec \xi }\,\,\,
{\rm and}\,\,\,
\delta k = - {\delta m^2 \over 2\om} {1
\over V^{(\nu)} - \vec V^{(\pi)} \vec \xi }
\label{deltaomfin}
\ee 
One sees that generally both $\delta \om$ and $\delta k$ are non-vanishing.
Only in the case of pion decay at rest, $\delta \om = 0$ but $\delta k$ is 
non-zero in any case. Substituting the obtained results into expression
(\ref{deltaphi}) for the phase difference we come to the standard 
expression~(\ref{deltaphist}) if $V_\pi = 0$. This result
shows a remarkable stability with respect
to assumptions made in its derivation. However if the 
pion is moving, then the oscillation phase contains an extra term 
\be
\delta \Phi = {\delta m^2 \over 2\om}\,\, {\vec \xi \left(\vec r -
\vec V^{(\pi)} t\right) \over V^{(\nu)} - \vec \xi\, \vec V^{(\pi)} }=
{r\,\delta m^2  \over 2 k } + {(\vec \xi\, \vec V^{(\pi)})
(r-V^{(\nu)} t ) \over
V^{(\nu)} - \vec \xi \vec V^{(\pi)} }
\label{deltaphivpi}
\ee
This extra term would lead to
a suppression of oscillation after averaging over 
time. This suppression is related to the motion of the source and reflects
the uncertainty in the position of pion at the moment of decay. So this
result can be understood in the frameworks of the standard naive 
approach.

Similar expression can be derived for the case when both neutrino and
muon from the decay $\pi \rar \mu +\nu_j$ are registered in the space-time
points $\vec r_\nu, t_\nu$ and $\vec r_\mu, t_\mu$ respectively. 
This case was considered in refs.~\cite{dolgov97,nauenberg99}. 
Here we will use the same approach as described above when
the muon is not registered. The only
difference is that in eq.~(\ref{ajn}) for the oscillation amplitude
we have to substitute the Green's 
function of muon $G_\mu (\vec r_\mu -r_s, t_\mu - t_s)$
instead of $\psi_n (\vec r_s,t_s)$. 
The calculations are essentially the same and after some
algebra the following expression for the oscillation amplitude
is obtained:
\bea
A_{\mu,\nu} &\sim& {V_\mu V_\nu \over r_\mu r _\nu}
\Theta ( L_\mu + L_\nu ) \exp \left[ -{\gamma (L_\mu + L_\nu) \over
2 (V_\mu + V_\nu) } \right]
\tilde C (V_\mu L_\nu - V_\nu L_\mu ) \nonumber \\
&& \exp {\left[ i \left( k^{(0)}_\mu r_\mu + k^{(0)}_\nu r_\nu - 
E^{(0)}_\mu t_\mu - E^{(0)}_\nu t_\nu \right)\right]}
\label{amunu}
\eea
where $L=Vt-r$. Each kinematic variable depends upon the neutrino state
$j$, so they should contain sub-index $j$. The upper indices ``0'' mean that
these momenta and energies are the central values of the corresponding wave
packets, so that the classical relation $\vec k^{(0} = \vec V E^{(0)}$  
holds for them. Here the direction of momenta 
are defined as above along the vector indicated to the observation point.
However, the kinematics in this case
is different from the previous one and the change of the energy and 
momentum of each particle for reactions with different sorts of 
neutrinos are related through the equations: 
\be
\delta E_\mu + \delta E_\nu =0\,\,\, {\rm and}\,\,\,
\delta \vec k_\mu + \delta \vec k_\nu =0.
\label{deltaek}
\ee 
For the central values of momenta the following relations are evidently true,
$\delta k_\nu = \delta E_\nu - \delta m^2 / 2 k_\nu$, and the similar one for
the muon without the last term proportional to the mass difference.
Correspondingly the phase of oscillation is given by the expression
\bea
\delta \Phi = \delta k_\mu r_\mu - \delta E_\mu t_\mu + \delta k_\nu r_\nu
- \delta E_\nu t_\nu = \nonumber \\
\delta E_\nu \left( {r_\mu -V_\mu t_\mu \over V_\mu} -
 {r_\nu -V_\nu t_\nu \over V_\nu} \right) -{\delta m^2 \over 2 k_\nu} \, r_\nu
\label{deltaphimunu}
\eea
The first two terms in the phase are proportional to the argument
of $\tilde C$ in eq.~(\ref{amunu}) and thus they give the contribution equal
to the size of the the source, i.e. to the wave packet of the initial pion.
If the latter is small (as usually the case) we obtain again the standard
expression for the oscillation phase.

\section{Matter effects}

Despite extremely weak interactions of neutrinos, matter may have a
significant influence on the oscillations if/because the mass difference
between the propagating eigenstates is very small.  
Description of neutrino oscillations in matter was first done in 
ref.~\cite{wolfenstein}. Somewhat later a very important effect of resonance
neutrino conversion was discovered~\cite{mikheev}, when even with a very 
small vacuum mixing angle, mixing in medium could reach the maximal value. 

Hamiltonian of free neutrinos in the mass eigenstate basis has the form:
\bea
{\cal H}_m^{(1,2)}= \left( \begin{array}{cc} E_1 & 0 \\
0 & E_2 \end{array} \right)
\label{hm}
\eea
where $E_j = \sqrt{ p^2 +m_j^2}$. In the interaction basis ${\cal H}_m$ is
rotated by the matrix~(\ref{u2}):
\bea
{\cal H}_m^{(a,b)} = U {\cal H}_m^{(1,2)} U^{-1} = 
 \left( \begin{array}{cc} \cos^2\theta\,E_1+ \sin^2\theta\, E_2  
&g \sint\cost \\
g\sint\cost  & \sin^2\theta\, E_1 +\cos^2\theta\, E_2 \end{array} \right)
\label{hmab}
\eea
Here $g=\delta m^2 /2E$ and we returned to the more usual notation $E$
for neutrino energy.
 
The interaction Hamiltonian is diagonal in the interaction basis and if
only first order effects in the Fermi coupling constant, $G_F$, are taken
into account, then the Hamiltonian can be expressed through refraction
index, $n_a$, of flavor $a$-neutrino in the medium (recall that the
deviation of refraction index from unity
is proportional to the forward scattering amplitude and
thus contains $G_F$ in the first power):
\bea
H_{int}^{(a,b)} =
 \left( \begin{array}{cc} E\,(n_a-1)  & 0 \\
0 & E\,(n_b - 1) \end{array} \right)
\label{hint}
\eea 
where a small difference between $E_1$ and $E_2$ in front of small 
factors $(n-1)$ was neglected.

Thus, up to a unit matrix, the total Hamiltonian in the interaction basis
can be written as
\bea
H_{tot}^{(a,b)} =
 \left( \begin{array}{cc} f & g\stw /2 \\
g\stw/2 & 0 \end{array} \right)
\label{htot}
\eea 
where $f=g\ctw + E \delta n$ and $\delta n = n_a -n_b$. 
This matrix is easy to diagonalize. Its eigenvalues are
\be
\lambda_{1,2} = { f \pm \sqrt{ f^2 + g^2 \sin^2 2\theta} \over 2}
\label{lambda12}
\ee
and the eigenstates in matter (up to normalization factor) are
\bea
|\nu_{1,2} \rangle = |\nu_a\rangle + {g \stw \over f\pm \sqrt{f^2+
g^2 \sin^2 2\theta}}\, |\nu_b\rangle
\label{nu12}
\eea

Refraction index may change with time, as happens in cosmology, or with
space point, if neutrinos propagate in inhomogeneous medium, for example
from the center of the Sun to its surface. If somewhere (or sometime) $f$
vanishes then the resonance transition of one neutrino species to another
is possible~\cite{mikheev}. Indeed let us assume that $\nue$ and $\num$
are mixed with a small vacuum mixing angle $\theta$  
and that initially an electronic neutrino was
produced in vacuum. So the initial propagating state 
would be mostly $\nue$:
\be
|\nu_1\rangle_{in} = |\nu_e \rangle + (1/2) \tan 2\theta
\,\, |\nu_\mu\rangle
\label{nu1in}
\ee
After propagation in the media where the function $f$ changes sign
passing through zero, the propagating state would become mostly $\num$:
\be
|\nu_1\rangle_{fin} = |\nu_e \rangle - (2 /\tan 2\theta)
\,\, |\nu_\mu\rangle
\label{nu1fin}
\ee
This effect of resonance conversion may play an important role in the
resolution of the solar neutrino problem and in cosmology.

\section{Neutrino oscillations in cosmology}
\subsection{A brief (and non-complete) review} 

Neutrino oscillations in the primeval plasma is significantly different from
e.g. solar neutrino oscillations in the following two important aspects.
First, cosmological plasma is almost charge symmetric. A relative
excess of any particles over antiparticles is believed to be at the level
$10^{-9}-10^{-10}$, while in stellar material the asymmetry is of
order unity. Neutrino oscillations in the early universe may change the
magnitude of asymmetry in the sector of active neutrinos. On the other
hand, the asymmetry has a strong influence on the oscillations through
the refraction index of the primeval plasma (see below). It leads to a
strong non-linearity of the problem and makes calculations quite complicated.
Second important point is that neutrino mean free path in the early universe
is quite small at high temperatures and hence breaking of coherence 
becomes essential. Because of that one cannot use wave functions for
description of oscillations and should turn to the density matrix formalism.
It leads to a great complexity of equations. 
Kinetic equations for density matrix with the account of neutrino
scattering and annihilation were derived in 
the papers~\cite{dolgov81,stodolsky87,raffelt93,sigl93}. In 
ref.~\cite{dolgov81}, where the impact of neutrino oscillations on 
big bang nucleosynthesis (BBN) was first considered, only the second order
effects, proportional to $G^2_F$, were taken into account,
while the deviation of refraction index 
from unity was neglected. This approximation is valid for a 
sufficiently high $\delta m^2$. In a subsequent paper~\cite{khlopov}
implications for BBN of possible CP-violating effects in oscillations
were discussed. Earlier works on neutrino oscillations also include
refs.~\cite{fargion84,langacker86,langacker87}. 
The calculations of refraction 
index in cosmological plasma were performed in ref.~\cite{notzold88}.
The results of this work permitted to make more accurate calculations of
the role played by neutrino oscillations in 
BBN~\cite{barbieri90}$^-$\cite{kostelecky96b}. 

It was noticed in ref.~\cite{barbieri91} that the oscillations between
an active and sterile neutrinos could generate an exponential rise of
lepton asymmetry in the sector of active neutrinos. The origin of this
instability is the following. Since lepton
asymmetry comes with the opposite sign to refraction indices of neutrinos
and antineutrinos (see below section~\ref{refr}),
it may happen that the transformation of antineutrinos
to their sterile partners would proceed faster than similar transformation
of neutrinos, especially if resonance conditions are fulfilled. 
It would lead to an increase of the asymmetry and through the refraction 
index to more favorable conditions for its rise. However it was 
concluded~\cite{barbieri91} that the rise is not significant and the
effect of the generated asymmetry on BBN is small. This conclusion
was reconsidered in ref.~\cite{foot96} where it was argued that asymmetry
generated by this mechanism could reach very large values, close to unity, 
and this effect,
in accordance to the earlier paper of the same group~\cite{foot95}, would
have a significant influence on primordial abundances. This result 
attracted a great attention and was confirmed in several subsequent
publications~\cite{foot97}$^-$\cite{dibari99c}. Moreover, some works
showed not only rising and large asymmetry but also a chaotic behavior of
its sign~\cite{shi96}$^-$\cite{sorri99}. 

Due to complexity of equations some simplifying
approximations were made in their solutions. First accurate
numerical solution of (almost) exact equations were done in 
refs.~\cite{kirilova97}$^-$\cite{kirilova99b}.
The most essential approximation was that the lost of coherence was 
described by the term $\gamma (\rho_{eq} - \rho)$ instead of the exact
two-dimensional 
collision integral taken from the square of the scattering amplitude over 
momenta. In fact this approximation was used in all the papers. 
The calculations of refs.~\cite{kirilova97}$^-$\cite{kirilova99b}  
were performed for rather small mass difference, 
$\delta m^2 < 10^{-7}\, {\rm eV}^2$. Neither
chaoticity, nor a considerable rise of the asymmetry were found. For
a larger mass difference a strong numerical instability was observed.
However this result is not in a contradiction with other papers because
the latter found the above mentioned effects for much larger mass
differences. In our work~\cite{dolgov99} we tried to extend the range of
validity of direct numerical calculations to large $\delta m^2$. We were
able to proceed only up to $\delta m^2 \approx 10^{-6}$. For higher
values we did not find a way to avoid numerical instability. To overcome the
problem we analytically transformed kinetic equations to a simpler form
that permitted a stable numerical integration. To this end an 
expansion in terms of a
large parameter that corresponds to a high frequency of oscillations was 
used. This method is rather similar to the well known separation of fast 
and slow variables in differential equations.
According to our results the asymmetry may rise, but only 
by 5-6 orders of magnitude (from initial $10^{-10}$), i.e. 4-5 orders of 
magnitude weaker than was obtained in other 
papers~\cite{foot96,foot97,bell98,foot99,dibari99a,dibari99b,dibari99c}.
However in agreement with these papers no chaoticity was found. 
Our results were criticized in ref.~\cite{sorri99} where it was argued that
lepton asymmetry generated by oscillations must be chaotic.
However the author misunderstood our calculations and criticized us for the
things that we never did and, second, the approximation used in that 
paper~\cite{sorri99}, namely calculations in terms of one fixed value
of neutrino momentum that was chosen to be equal to the thermally 
averaged one,
is intrinsically inappropriate for the solution of the problem of chaoticity.
In that approximation even neutrino oscillations in vacuum would result
in chaotic lepton asymmetry.

Thus, most probably a chaotic amplification of the lepton
asymmetry does not take place,
however the exact magnitude of the amplification remains uncertain. As 
discussed in ref.~\cite{foot99,dibari99c} their recent calculations
are exact and do not suffer from numerical instability. On the other hand,
we do not see any shortcomings of our 
semi-analytical approach~\cite{dolgov99}, all
approximations are well under control and numerical part of calculations 
is quite simple and stable. More work is necessary to resolve the
contradictions both in magnitude and possible chaoticity.

\subsection{Refraction index \label{refr}}

In this section we derive the Schroedinger equation for neutrino wave
function in the primeval plasma. We will start with the neutrino
quantum operator $\nu_a (x)$ of flavor $a$ that satisfies the usual 
Heisenberg equation of motion:
\bea
\left( i \ds - {\cal M} \right) \nu_a (x)
+ {g \over 2\sqrt 2}\,\delta_{ae}\, W_\alpha (x) O_\alpha^{(+)} e (x)
+{ g \over 4\cos \theta_W} Z_\alpha (x) O_\alpha^{(+)} \nu_a (x) =0
\label{dirac}
\eea
where ${\cal M}$ is the neutrino mass matrix, $W(x)$, $Z(x)$ and $e(x)$ 
are respectively the quantum operators of intermediate bosons and electrons,
and  $O_\alpha^\pm = \gamma_\alpha \left( 1\pm \gamma_5\right)$.
We assumed that the temperature of the plasma is in MeV range and thus
only electrons are present in the plasma. 

Equations of motion for the field operators of $W$ and $Z$ bosons have the
form
\be 
G_{W,\alpha\beta}^{-1} W_\beta (x) = 
{ g\over 2\sqrt 2} \bar \nu_a (x) O^{(+)}_\alpha \nu_a (x)\,,
\label{gw}
\ee
\bea 
G_{Z,\alpha\beta}^{-1} Z_\beta (x) = 
{g\over 4\cos \theta_W}\left[ \bar \nu_a (x) O_\alpha^{(+)} \nu_a (x) + 
\right .
\nonumber \\
\left .
\left( 2 \sin^2 \theta_W -1\right) \bar e(x) O_\alpha^{(+)} e (x)
+2\sin^2 \theta_W \, \bar e(x) O_\alpha^{(-)} e (x) \right]
\label{gz}
\eea
where
the differential operators $G^{-1}_{W,Z}$ are inverse Green's functions
of $W$ and $Z$. In momentum representation they can be written as
\be
G_{\alpha\beta} = {g_{\alpha\beta} - q_\alpha q_\beta / m^2 \over
m^2 - q^2 }
\label{g}
\ee
It can be be shown that the term $q_\alpha q_\beta/m^2$ gives
contribution proportional to lepton masses and can be neglected. 

In the limit of small momenta, $q\ll m_{W,Z}$, equation (\ref{gw}) can be
solved as
\be
W_\alpha (x) = - {g \over 2\sqrt 2 \, m^2_W} \left( 1 -
{\partial^2 \over m^2_W} \right) \left( \bar e(x) 
O_\alpha^{(+)} \nu_e (x)\right)
\label{wx}
\ee
A similar expression with an evident substitution for the r.h.s. can be 
obtained for $Z_\alpha (x)$. These expressions should be substituted
into eq.~(\ref{dirac}) to obtain equation that contains only field operators
of leptons, $\nu_a (x)$ and $e(x)$:
\bea
\left( i\ds - \cal M \right) \nu_a (x) =
{G_F \over \sqrt 2} \left\{ \delta_{ae} \left[\left( 1-{\partial^2 
\over m_W^2}\right) \left(\bar e(x) O_\alpha^{(+)} \nu_e (x) \right) \right]
O_\alpha^{(+)} e(x)  + \right . \nonumber \\
\left .
{1\over 2} \left[\left( 1-{\partial^2  \over m_Z^2}\right) 
\left( \bar \nu_b(x) O_\alpha^{(+)} \nu_b (x) + 
\left( 2\sin^2\theta_W -1\right)
\bar e(x) O_\alpha^{(+)} e(x) + 
 \right . \right. \right . \nonumber \\
\left . \left . \left .
2\sin^2\theta_W\,\, \bar e(x) O_\alpha^{(-)} e(x)
\right)\right] O^{(+)}_\alpha \nu_a (x) \right\}
\label{dsnua}
\eea

Neutrino wave function in the medium is defined as      
\be
\Psi_a (x) = \langle A | \nu_a (x) |A + \nu^{(k)} \rangle
\label{psinu}
\ee
where $A$ describes the state of the medium and $\nu^{(k)}$ is a 
certain one-neutrino state, specified by quantum numbers $k$,
e.g. neutrino with momentum $\vec k$.  
Equation of motion for this wave function can be found from 
eq.~(\ref{dsnua}) after averaging over medium. 
The theory is quantized perturbatively in the standard way. We define
the free neutrino operator $\nu_a^{(0)}$ that satisfies the
equation of motion:
\be
\left ( i \ds - \cal M \right ) \nu_a^{(0)} (x) = 0
\label{nufree}
\ee
This operator is expanded as usually
in terms of creation-annihilation operators: 
\bea
\nu^{(0)} (x) = \int {d^3 k \over (2\pi)^3 \sqrt{ 2 E_k} }
\sum_s \left( a_k^s u^s (k) e^{ikx} + b_k^{s\,\dagger} v^s (k) e^{ikx}
\right)
\label{nu0x}
\eea 
and one-particle state is defined
as $|\nu^{(k)}\rangle = a^\dagger_k | {\rm vac} \rangle$. 

The equation of motion for the neutrino wave function $\Psi_a (x)$ can be
obtained from expression~(\ref{psinu}) perturbatively by applying 
the operator $\left ( i \ds - \cal M \right )$ and using eq.~(\ref{dsnua}) 
with free neutrino operators $\nu_a^{(0)}$ in the r.h.s. After some
algebra which mostly consists in using equations of motion for the free
fermion operators and (anti)commutation relations between the 
creation/annihilation operators, one would obtain the equation of the form:
\be
i\partial_t \Psi (t) = \left( {\cal H}_m+ V_{eff} \right) \Psi  
\label{dtpi}
\ee
where ${\cal H}_m$ is the free Hamiltonian; in the mass eigenstate
basis it has the form 
$ {\cal H}_0 = {\rm diag} \left[ \sqrt{p^2 + m^2_j}\,\,\right]$.
The matrix-potential $V_{eff}$
describes interactions of neutrinos with media and  
is diagonal in the flavor basis. Up to the factor
$E$ (i.e. neutrino energy) it is essentially the refraction index of neutrino
in the medium. The potential contains two terms. The first one comes
from the averaging of the external current $J\sim \bar l O_\alpha l$.
Due to homogeneity and isotropy of the plasma only its time component
is non-vanishing and proportional to the charge asymmetry
(i.e. to the excess of particles over antiparticles) in the plasma. This
term has different signs for neutrinos and antineutrinos. However
the interactions of neutrinos with the medium is not always of the 
(current)$\times$(current) form due to non-locality related to the
exchange of $W$ or $Z$ bosons. If incoming and outgoing neutrinos interact
in different space-time points, the interaction with the medium cannot
be written as an interaction with the external current. The contribution of 
such terms is inversely proportional to $m^2_{W,Z}$ but formally it is of
the first order in $G_F$. With these two types of contributions 
the diagonal matrix elements of the
effective potential for the neutrino of flavor $a$ has the form:
\be
V_{eff}^a =
\pm C_1 \eta G_FT^3 + C_2^a \frac{G^2_F T^4 E}{\alpha} ~,
\label{nref}
\ee
where $E$ is the neutrino energy, $T$ is the temperature of the
plasma, $G_F=1.166\cdot 10^{-5}$ GeV$^{-2}$ is the Fermi coupling
constant, $\alpha=1/137$ is the fine structure constant, and the signs
``$\pm$'' refer to anti-neutrinos and neutrinos respectively (this
choice of sign describes the helicity state, negative for $\nu$ and
positive for $\bar\nu$). According to ref.~\cite{notzold88} the
coefficients $C_j$ are: $C_1 \approx 0.95$, $C_2^e \approx 0.61$ and
$C_2^{\mu,\tau} \approx 0.17$.  These values are true in the limit of
thermal equilibrium, but otherwise these coefficients are some
integrals from the distribution functions over momenta. 
For oscillating neutrinos
deviations from thermal equilibrium could be significant and in this case
expression (\ref{nref}) should be modified. However it is technically
rather difficult to take this effect into account
in numerical calculations and the simplified version (\ref{nref}) is used.

\subsection{Loss of Coherence and Density Matrix \label{coher}}

Breaking of coherence appears in the second order in the Fermi coupling
constant $G_F$, so that equations of motion for the operators of
{\it all} leptonic fields (including electrons) should be solved up to
the second order in $G_F$. Since the calculations are quite lengthy, 
we only sketch the derivation here. In the considered
approximation the lepton operators $l(x)$, where $l$ stands for neutrino
or electron, in the r.h.s. of eq.~(\ref{dsnua}) 
should be expanded up to the first 
order in $G_F$. The corresponding expressions
can be obtained from the formal solution of eq.~(\ref{dsnua}) up to first
order in $G_F$. Their typical form is the following:
\be
l = l_0  +  G_l * ({\rm r.h.s.}_0) 
\label{l1}
\ee
where the matrix (in neutrino space) $G_l$ is the Green's function of the
corresponding lepton and ${\rm r.h.s.}_0$ is the right hand side of 
eq.~(\ref{dsnua}) taken in the lowest order in $G_F$, i.e with
lepton operators taken in the zeroth order, $l=l_0$. 
Expression~(\ref{l1}) should be substituted back into eq.~(\ref{dsnua})
and this defines the r.h.s. up to the second order in $G_F$ in terms of
free lepton operators $l_0$.
Of course in the second approximation we neglect the non-local terms, 
$\sim 1/m_{W,Z}^2$. 

Now we can derive kinetic equation for the density matrix of neutrinos,
$\hat {\rho}^i_j = \nu^i \nu_j^*$, where over-hut indicates that 
$\hat \rho$ is a quantum operator. The $C$-valued density matrix is
obtained from it by taking matrix element over the medium,
$\rho = \langle \hat \rho \rangle$. We should apply to it the differential 
operator $(i\ds - {\cal M})$ and use eq.~(\ref{dsnua}). 
The calculations of matrix elements of the free lepton operators 
$l_0$ are straightforward and can be achieved by using the standard 
commutation relations. There is an important difference between equations
for the density matrix and the wave function. The latter contains only terms
proportional to the wave function, $i\partial_t \Psi = {\cal H} \Psi$,
while equation for density matrix contains source term that does not
vanish when $\rho = 0$. Neutrino production or destruction is described by
the imaginary part of the effective Hamiltonian. The latter is not hermitian
because the system is not closed. By the optical theorem the imaginary part
of the Hamiltonian is expressed through the cross-section of neutrino 
creation or annihilation. Such terms in kinetic equation for the
density matrix are similar to the ``normal'' kinetic equation for the 
distribution functions:
\be
{d f_1 \over dt} = I_{coll}
\label{dfdt}
\ee
where the collision integral is the integral over the proper phase space
from the following combination of the distribution functions 
\be
F= -f_1 f_2 (1-f_3)(1-f_4) + f_4 f_3 (1-f_1)(1-f_2)
\label{Ff}
\ee
A similar combination appears for the case of oscillating neutrinos but
with a rather complicated matrix structure. For example there can be
the contribution of the form of the anti-commutator:
\be
-\left\{ \rho_1, g (1-\rho_3) g \rho_2 (1-\rho_4) g \right \}
+ \left({\rm inverse\,\,\, reaction} \right)
\label{rhorho}
\ee
and a few more of different structure that results in a very lengthy
expression. The matrix $g$ describes neutrino
interactions. It is diagonal in the flavor basis and has entries proportional
to matrix elements squared of the relevant reactions. 
The Fermi-blocking factors $(1-\rho)$ appear when one takes matrix elements
of neutrino operators over the medium in which neutrino occupation
numbers may be non-zero.

We will not present here the complete form of the equation. It can be found
e.g. in the paper~\cite{sigl93}. Moreover in all the applications a 
``poor man'' substitution has been done: all terms describing neutrino
production or destruction were mimicked by
\be
- \Gamma \left(\rho - \rho_{eq} \right)
\label{gammarho}
\ee
where $\rho_{eq}$ is the equilibrium value of the density matrix, i.e.
the unit matrix multiplied by the equilibrium distribution function
\be
f_{eq} = \left[ \exp ( E/T  - \xi ) +1 \right]
\label{feq}
\ee  
and the coefficient $\Gamma$ is the reaction rate (see below).

\subsection{Oscillations and lepton asymmetry \label{leptas}}

As we have already mentioned oscillations between active and sterile
neutrinos may induce a significant lepton asymmetry in the sector of active
neutrinos. Now we will consider this phenomenon in some more detail.
Basic equations governing the evolution of the density matrix are:
\bea
i(\partial_t -Hp\partial_p) \rho_{aa}
&=& F_0(\rho_{sa}-\rho_{as})/2 -i \Gamma_0 (\rho_{aa}-f_{eq})~,
\label{dotrhoaa} \\
i(\partial_t -Hp\partial_p)  \rho_{ss}
&=& -F_0(\rho_{sa}-\rho_{as})/2~,
\label{dotrhoss} \\
i(\partial_t -Hp\partial_p) \rho_{as} &=&
W_0\rho_{as} +F_0(\rho_{ss}-\rho_{aa})/2-
i\Gamma_1 \rho_{as} ~,
\label{dotrhoas}\\
i(\partial_t -Hp\partial_p) \rho_{sa} &=& -W_0\rho_{sa} -
F_0(\rho_{ss}-\rho_{aa})/2- i\Gamma_1 \rho_{sa}  ~,
\label{dotrhosa}
\eea
where $a$ and $s$ mean ``active'' and ``sterile'' respectively, $F_0=\dm
\sin 2\theta / 2E$, $W_0= \dm\cos 2\theta /2E + V_{eff}^a$,
$H=\sqrt{8\pi \rho_{tot}/3M_p^2}$ is the Hubble parameter, $p$ is the
neutrino momentum.
The antineutrino density matrix satisfies the similar set of equations
with the opposite sign of the antisymmetric term in $V_{eff}^a$ and with
a slightly different damping factor $\bar\gamma$ (this difference is
proportional to the lepton asymmetry in the primeval plasma).

Equations (\ref{dotrhoaa}-\ref{dotrhosa})
account exactly for the first order terms described by
the refraction index, while the second order terms describing breaking of
coherence are approximately modeled by the damping coefficients $\Gamma_j$
in accordance with eq.~(\ref{gammarho}).
The latter are equal to~\cite{harris}:
\be
\Gamma_0 = 2\Gamma_1  = g_a \frac{180 \zeta(3)}{7 \pi ^4}
\, G_F^2 T^4 p  ~.
\label{gammaj1}
\ee
In general the coefficient $g_a(p)$ is a momentum-dependent
function, but in the approximation of neglecting $[1-f]$ factors in the
collision integral it becomes a constant~\cite{bell99} equal respectively
to $g_{\nu_e} \simeq 4$ and $g_{\nu_\mu,\mu_\tau} \simeq
2.9$ \cite{enqvist92a}.  In the following we will use more accurate values
found from the thermal average of the complete
electro-weak rates (with factors $[1-f]$ included), which we
calculated numerically from our Standard Model code
\cite{dolgov97}. This gives us $g_{\nu_e} \simeq 3.56$
and $g_{\nu_\mu,\mu_\tau} \simeq 2.5$.

It is convenient to introduce new variables
\be
x=m_0 R(t)\,\, {\rm and}\,\,   y=p R(t)
\label{xy}
\ee
where $R(t)$ is the cosmological scale factor so that $H=\dot
R/R$ and $m_0$ is an arbitrary mass (just normalization), we choose
$m_0 =1$ MeV. In the approximation that we will work, we assume that
$\dot T =-HT$, so that we can take $R=1/T$. In terms of these
variables the differential operator $(\partial_t -Hp\partial_p)$
transforms to $Hx\partial_x$. We will normalize the density matrix
elements to the equilibrium function $f_{eq}$:
\be
\rho_{aa} &=& f_{eq}(y) [1+a(x,y)],\,\, \rho_{ss} = f_{eq}(y) [1+s(x,y)]~, \\
\label{rhoaa}
\rho_{as} &=& \rho_{sa}^* = f_{eq}(y)[h(x,y)+i l(x,y)]~,
\label{hil}
\ee
and express the neutrino mass difference $\delta m^2$ in eV$^2$.

As the next step we will take the sum and difference of
eqs.~(\ref{dotrhoaa})-(\ref{dotrhosa}) for $\nu$ and $\bar\nu$. The
corresponding equations have the following form:
\be
s_{\pm}' &=& F l_{\pm}~, \label{firstasyms} \\
a_{\pm}' &=& - F l_{\pm} - 2 \gamma_+ a_{\pm} - 2 \gamma_- a_{\mp} ~,\\
h_{\pm}' &=& U l_{\pm} - V Z l_{\mp} - \gamma_+ h_{\pm} -
\gamma_- h_{\mp} ~,\\
l_{\pm}' &=&\frac{F}{2}(a_{\pm} - s_{\pm}) - U h_{\pm} + V Z h_{\mp} -
\gamma _+ l _{\pm}
- \gamma_- l_{\mp}~,
\label{firstasyml}
\ee
where $a_{\pm} = (a \pm {\bar a})/2$ etc, and the prime means
differentiation with respect to $x$.  We have used $W=U \pm VZ$,
$\gamma =\Gamma_1/Hx$, and $\gamma_{\pm} =(\gamma \pm \bar\gamma)/2$,
where $\gamma_-$ parameterizes the difference of interaction rates
between neutrino and anti-neutrinos, which is proportional to the
neutrino asymmetry. With the approximation
$\rho_{tot} \simeq 10.75
\pi^2 T^4/30$, the expressions for $U$, $V$, and $Z$ become:
\be
U &=& 1.12\cdot 10^9\cos 2\theta \delta m^2 {x^2\over y}+26.2{y\over x^4},\\
V &=&\frac{29.6}{x^2},\\
Z &=& 10^{10}\left (\eta_{o}
- \int \frac{dy}{4 \pi^2} ~y^2 f_{eq} ~a_-\right )~,
\label{VandZ}
\ee
where $\eta_o$ is the asymmetry of the other particle species:
\be
\eta^e_o &=& 
2\eta_{\nue} +\eta_{\num} + \eta_{\nut} +\eta_{e}-\eta_{n}/2 \,\,\,\,\,
 ( {\rm for} \,\, \nue)~,
\label{etanue} \\
\eta^\mu_o &=& 
2\eta_{\num} +\eta_{\nue} + \eta_{\nut} - \eta_{n}/2\,\,\,\,\,
({\rm for} \,\, \num)~,
\label{etanumu}
\ee
and $\eta$ for $\nut$ is obtained from eq.~(\ref{etanumu}) by the interchange
$\mu \lrar \tau$.
The asymmetry is normalized in the same way as the
neutrino asymmetry (the second term in (\ref{VandZ})). Here we have
implicitly assumed that $\nu_a=\nu_e$.

Up to this point our equations are essentially the same as those used by 
other groups. The equations look rather innocent and at first sight one
does not anticipate any problem with their numerical solution. 
However the contribution from $Z$ could be quite large with the
increasing magnitude of the asymmetry. The exact value of the latter is
determined by a delicate cancellation of the contributions from all
energy spectrum of neutrinos. The function $a_-$ under momentum integral
is quickly oscillating and very good precision is necessary to calculate
the integral with a desired accuracy. Even  a small numerical error
results in a large instability. To avoid this difficulty we analytically
separated fast and slow variables in the problem and reduced this
set of equations to a single differential equation for the asymmetry
that can be easily numerically integrated. The corresponding
algebra is somewhat complicated and we will not discuss it here. One can
find the details in ref.~\cite{dolgov99}. 

We have found that asymmetry practically does not rise for a large mixing
angles, $\sin \theta > 0.01$. For smaller mixings some rise of the
asymmetry is observed, though much weaker than that obtained in 
refs.~\cite{foot96,foot97,bell98,foot99,dibari99a,dibari99b,dibari99c}.
For example for $\delta m^2 = 1$ eV$^2$ we found that asymmetry
rises by approximately 5 orders of magnitude reaching the value around 
$10^{-5}$ for $\sin 2\theta$ in the
interval $10^{-5}-10^{-3}$. For $\delta m^2 = 10^6$ eV$^2$ the symmetry
could rise up to 0.01 for $\sin 2\theta = 3\cdot 10^{-6} -
 3\cdot 10^{-5}$, and only for huge mass difference $\delta m^2 = 10^9$
the asymmetry may reach unity in a rather narrow range of mixing angles,
$\sin 2\theta = 2\cdot 10^{-6} - 4\cdot 10^{-6}$. For $\sin 2\theta$
outside the indicated limits the asymmetry does not rise.
For even larger mass difference, $\delta m^2 = 10^{12}$ eV$^2$,
the asymmetry practically does not rise. 

Resolution of the contradiction between different groups is very important
for the derivation of the constraints on the parameters of the oscillations
from big bang nucleosynthesis (BBN). There are several effects by which the 
oscillations may influence abundances of light elements:
\begin{enumerate}
\item{}
If sterile neutrinos are excited by the 
oscillations then the effective number
of neutrino species at nucleosynthesis would be larger than 3. This effect,
as is well known, results in an increase of mass fraction of helium-4
and deuterium. 
\item{}
Oscillations may distort the spectrum of neutrinos and, in particular, of
electronic neutrinos. The sign of the effect is different, depending on 
the form of spectral distortion. A deficit of electronic neutrinos at
high energy results in a smaller mass fraction of helium-4, while a deficit
of $\nue$ at low energy works in the opposite direction. A decrease
of total number/energy density of $\nue$ would result in an earlier 
freezing of neutrino-to-proton ratio and in a larger fraction of helium-4.
\item{}
Oscillations may create an asymmetry between $\nue$ and anti-$\nue$.
The $n/p$ ratio in this case would change as $n/p \sim \exp (\mu/T)$,
where $\mu$ is the chemical potential corresponding to the asymmetry. 
With the present day accuracy of the data the asymmetry in the sector
of electronic neutrinos could be at the level of a few per cent, i.e.
much larger than the standard $10^{-10}$. Even if asymmetry is strongly
amplified but still remains below 0.01 its direct influence on BBN would
be negligible. It may however have an impact on the nucleosynthesis in an
indirect way. Namely, the rise of the asymmetry by several orders of
magnitude could suppress neutrino oscillations through refraction index
so that new neutrino species corresponding to sterile neutrinos are not 
efficiently excited. 
\end{enumerate}

Thus one sees that the effects of oscillations may result both in 
a reduction or in an increase the effective number of neutrino species. 
In particular, oscillations may open room for additional particles 
at BBN. This conclusion strongly depends upon the magnitude of lepton
asymmetry generated by the oscillations. Thus a resolution of the 
controversies between different theoretical calculations would be
very important.

\section*{Acknowledgments} 
This work was partly supported by Danmarks Grundforskningsfond through its 
funding of the Theoretical Astrophysics Center. 
 
%
%
\nc{\advp}[3]{{\it  Adv.\ in\ Phys.\ }{{\bf #1} {(#2)} {#3}}}
\nc{\annp}[3]{{\it  Ann.\ Phys.\ (N.Y.)\ }{{\bf #1} {(#2)} {#3}}}
\nc{\apl}[3] {{\it  Appl. Phys. Lett. }{{\bf #1} {(#2)} {#3}}}
\nc{\apj}[3] {{\it  Ap.\ J.\ }{{\bf #1} {(#2)} {#3}}}
\nc{\apjl}[3]{{\it  Ap.\ J.\ Lett.\ }{{\bf #1} {(#2)} {#3}}}
\nc{\app}[3] {{\it  Astropart.\ Phys.\ }{{\bf #1} {(#2)} {#3}}}
\nc{\cmp}[3] {{\it  Comm.\ Math.\ Phys.\ }{{ \bf #1} {(#2)} {#3}}}
\nc{\cqg}[3] {{\it  Class.\ Quant.\ Grav.\ }{{\bf #1} {(#2)} {#3}}}
\nc{\epl}[3] {{\it  Europhys.\ Lett.\ }{{\bf #1} {(#2)} {#3}}}
\nc{\ijmp}[3]{{\it  Int.\ J.\ Mod.\ Phys.\ }{{\bf #1} {(#2)} {#3}}}
\nc{\ijtp}[3]{{\it  Int.\ J.\ Theor.\ Phys.\ }{{\bf #1} {(#2)} {#3}}}
\nc{\jmp}[3] {{\it  J.\ Math.\ Phys.\ }{{ \bf #1} {(#2)} {#3}}}
\nc{\jpa}[3] {{\it  J.\ Phys.\ A\ }{{\bf #1} {(#2)} {#3}}}
\nc{\jpc}[3] {{\it  J.\ Phys.\ C\ }{{\bf #1} {(#2)} {#3}}}
\nc{\jap}[3] {{\it  J.\ Appl.\ Phys.\ }{{\bf #1} {(#2)} {#3}}}
\nc{\jpsj}[3]{{\it  J.\ Phys.\ Soc.\ Japan\ }{{\bf #1} {(#2)} {#3}}}
\nc{\lmp}[3] {{\it  Lett.\ Math.\ Phys.\ }{{\bf #1} {(#2)} {#3}}}
\nc{\mpl}[3] {{\it  Mod.\ Phys.\ Lett.\ }{{\bf #1} {(#2)} {#3}}}
\nc{\ncim}[3]{{\it  Nuov.\ Cim.\ }{{\bf #1} {(#2)} {#3}}}
\nc{\np}[3]  {{\it  Nucl.\ Phys.\ }{{\bf #1} {(#2)} {#3}}}
\nc{\pr}[3]  {{\it  Phys.\ Rev.\ }{{\bf #1} {(#2)} {#3}}}
\nc{\pra}[3] {{\it  Phys.\ Rev.\ A\ }{{\bf #1} {(#2)} {#3}}}
\nc{\prb}[3] {{\it  Phys.\ Rev.\ B\ }{{{\bf #1} {(#2)} {#3}}}}
\nc{\prc}[3] {{\it  Phys.\ Rev.\ C\ }{{\bf #1} {(#2)} {#3}}}
\nc{\prd}[3] {{\it  Phys.\ Rev.\ D\ }{{\bf #1} {(#2)} {#3}}}
\nc{\prl}[3] {{\it  Phys.\ Rev.\ Lett.\ }{{\bf #1} {(#2)} {#3}}}
\nc{\pl}[3]  {{\it  Phys.\ Lett.\ }{{\bf #1} {(#2)} {#3}}}
\nc{\prep}[3]{{\it  Phys.\ Rep.\ }{{\bf #1} {(#2)} {#3}}}
\nc{\prsl}[3]{{\it  Proc.\ R.\ Soc.\ London\ }{{\bf #1} {(#2)} {#3}}}
\nc{\ptp}[3] {{\it  Prog.\ Theor.\ Phys.\ }{{\bf #1} {(#2)} {#3}}}
\nc{\ptps}[3]{{\it  Prog\ Theor.\ Phys.\ suppl.\ }{{\bf #1} {(#2)} {#3}}}
\nc{\physa}[3]{{\it Physica\ A\ }{{\bf #1} {(#2)} {#3}}}
\nc{\physb}[3]{{\it Physica\ B\ }{{\bf #1} {(#2)} {#3}}}
\nc{\phys}[3]{{\it  Physica\ }{{\bf #1} {(#2)} {#3}}}
\nc{\rmp}[3] {{\it  Rev.\ Mod.\ Phys.\ }{{\bf #1} {(#2)} {#3}}}
\nc{\rpp}[3] {{\it  Rep.\ Prog.\ Phys.\ }{{\bf #1} {(#2)} {#3}}}
\nc{\sjnp}[3]{{\it  Sov.\ J.\ Nucl.\ Phys.\ }{{\bf #1} {(#2)} {#3}}}
\nc{\sjp}[3] {{\it  Sov.\ J.\ Phys.\ }{{\bf #1} {(#2)} {#3}}}
\nc{\spjetp}[3]{{\it Sov.\ Phys.\ JETP\ }{{\bf #1} {(#2)} {#3}}}
\nc{\yf}[3]  {{\it  Yad.\ Fiz.\ }{{\bf #1} {(#2)} {#3}}}
\nc{\zetp}[3]{{\it  Zh.\ Eksp.\ Teor.\ Fiz.\ }{{\bf #1} {(#2)} {#3}}}
\nc{\zp}[3]  {{\it  Z.\ Phys.\ }{{\bf #1} {(#2)} {#3}}}
\nc{\ibid}[3]{{\sl  ibid.\ }{{\bf #1} {#2} {#3}}}
%
%

\end{document}